\documentclass[a4paper,11pt,twosided]{article}
\usepackage{graphicx,floatflt,amssymb,epsfig,amsmath}
\usepackage{pdfpages}
\usepackage{slashed,setspace,fancyhdr}
\begin{document}


\begin{center}
{\Large \bf Lepton flavour violating decay of 125 GeV Higgs boson to $\mu\tau$
channel and excess in $t\bar t H$}\\
\vskip 0.3cm
Biplob Bhattacherjee $^{a}$\footnote{Email: biplob@chep.iisc.ernet.in}, 
Sabyasachi Chakraborty $^{b}$\footnote{Email: tpsc3@iacs.res.in}, 
Swagata Mukherjee $^{c}$\footnote{Email: physics.swagata@gmail.com}
\vskip 0.3cm
{$^a$ Centre for High Energy Physics, Indian Institute of Science, Bangalore 560012, India}\\
{$^b$ Indian Association for the Cultivation of Science, Kolkata 700032, India}\\
{$^c$ Saha Institute of Nuclear Physics, Kolkata 700064, India} \\

\end{center}

\begin{abstract}
A recent search for the lepton flavor violating (LFV) decays of the Higgs boson, 
performed by CMS collaboration, reports an interesting deviation from the standard 
model (SM). The search conducted in the channel $H\rightarrow \mu\tau_e$ and 
$H\rightarrow \mu\tau_{\textrm{had}}$ shows an excess of $2.4\sigma$ signal events 
with 19.7 fb$^{-1}$ data at a center-of-mass energy $\sqrt s=8$ TeV. On the other 
hand, a search performed by CMS collaboration for the SM Higgs boson produced in 
association with a top quark pair ($t\bar t H$) also showed an excess in the same-sign 
di-muon final state. In this work we try to find out if these two seemingly uncorrelated
excesses are related or not. Our analysis reveals that a lepton flavour violating
Higgs decay ($H\rightarrow\mu\tau$) can partially explain the excess in the same
sign di-muon final state in the $t\bar t H$ search, infact brings down the excess
well within 2$\sigma$ error of the SM expectation. Probing such non-standard Higgs boson 
decay is of interest and might contain hints of new physics at the electroweak scale.

\end{abstract}



\section{Introduction}
The Higgs boson was hypothesised~\cite{Englert:1964et,Higgs:1964ia,Higgs:1964pj,
Guralnik:1964eu} in the year 1964 and since then experimental searches for this 
elusive boson have been performed in different collider experiments worldwide. 
Finally, after around 50 years of its theoretical proposition, a Higgs boson is 
discovered by both the ATLAS and CMS experiments of LHC at CERN, Geneva; the 
announcement of which was made on July 4, 2012~\cite{Aad:2012tfa,Chatrchyan:2012ufa}. 
With accumulation of more data throughout the year 2012, the properties 
of this newly observed boson was measured more accurately. The spin and 
parity properties and couplings of Higgs boson with fermions and bosons, 
the so-called $\kappa_f$ and $\kappa_v$, have been measured in different decay 
channels~\cite{Chatrchyan:2013mxa,Chatrchyan:2013iaa,Khachatryan:2014jba,
Aad:2015zhl,conf} of Higgs boson. Though most of the measurements are 
consistent with SM predictions within uncertainties, there is still some 
room for new physics beyond the SM, given the amount of uncertainties are 
still quite large in some cases.
\section{An excess in lepton flavour violating decay channel of Higgs boson}
In the framework of the SM, the interaction between
the Higgs boson and the SM fermions, written in the mass eigenstate 
basis is
\begin{equation}
\mathcal{L}_Y = -Y_{ij} \bar{f}_{L}^{i} f_{R}^{j} H + \textrm{h.c.}
\end{equation}
where $f_L$'s are the left handed fermion (lepton or quark) doublet and $f_R$'s 
denote the right handed fermion singlet. $Y_{ij}$ represent the Yukawa couplings 
in the mass basis and are diagonal in the paradigm of SM. In such a scenario, 
lepton flavor violating (LFV) Higgs decays are prohibited. However, such LFV 
decays can be incorporated in many beyond the SM (BSM) scenarios. For example, 
models with additional Higgs or other scalar fields~\cite{Sierra,Crivellin,Tobe,
Dipankar,Lee}, Higgs portals~\cite{deLima}, flavor symmetric models broken at 
the electroweak scale~\cite{Campos,Werner} and horizontal gauge symmetries~
\cite{Heeck} can explain such exotic decays of the Higgs boson. On the other 
hand, prototypical supersymmetric models with~\cite{Brignole,Ghosh} and without 
R-parity violation~\cite{Arhrib} are rather unlikely to explain such signals. 
However, in the context of this work, we will consider a model independent 
approach and will remain completely agnostic about different models which can 
give rise to LFV Higgs decays, more precisely $H\rightarrow \mu\tau$ channel. 
From the perspective of effective field theory (EFT), one can write down higher 
dimensional operators ($d>4$) suppressed by a new physics scale $\Lambda^{d-4}$. 
These operators would get generated by integrating out the `{\it non-standard}' 
heavy degrees of freedom. All of these operators would certainly leave a fingerprint 
in the low energy theory. As discussed in ref~\cite{Harnik:2012pb} a dimension 
6 operator of the form $\lambda_{ij}(\bar{f_L^i} f_R^j)H (H^{\dagger}H)/\Lambda^2 $
can indeed generate a non-diagonal Yukawa matrix $Y_{ij}$ in the mass eigenstate
basis of the fermions, which leads to LFV decays of the Higgs boson.
\par On the experimental frontier, the first direct search for lepton 
flavor violating decays of a Higgs boson to a muon-tau pair has been performed 
by the CMS collaboration~\cite{Khachatryan:2015kon}. The search has been conducted in 
two channels $H\rightarrow \mu\tau_e$ and $H\rightarrow \mu \tau_{had}$, 
where $\tau_{\textrm{had}}$ and $\tau_e$ represents hadronic and electronic
decays of $\tau$ respectively. 
A slight excess of signal events with a significance of 2.4$\sigma$ is observed,
which corresponds to a local p-value of 0.010\footnote{p-value is defined as the 
probability, under the background-only hypothesis (b), to obtain a value $q_0$ 
which is at least as large as that observed in data, $q_0^{data}$ : p-value = 
Prob $(q_0 \ge q_0^{data} | b)$. p-value measures how likely it is to get a 
certain experimental result as a matter of chance rather than due to a real effect.}.
A constraint on $BR(H \rightarrow \mu\tau)<1.51$\% at 95\% confidence level 
is set and the best fit branching fraction is $BR(H \rightarrow \mu\tau)
=(0.84^{+0.40}_{-0.37})$\%, as obtained by CMS. This limit on the 
branching ratio can be subsequently translated to 
constrain the Yukawa coupling $Y_{\mu\tau}$~\cite{Khachatryan:2015kon}.
Similarly, the ATLAS collaboration also looked into LFV decay modes of
the Higgs boson with 8 TeV centre-of-mass energy and with 20.3 fb$^{-1}$
data. The limit on the branching fraction $\textrm{Br} (H\rightarrow\mu\tau)$
is set at 1.85\%~\cite{Aad:2015gha}.
\vskip -3cm
\section{Another excess in $t\bar{t}H$ channel} \label{sec:tth}
Recently, a search for the SM Higgs boson produced in association with a 
top-quark pair ($t\bar{t}H$) is performed by the CMS 
collaboration~\cite{Khachatryan:2014qaa}. The search have been performed in 
different final state combinations, such as $\gamma\gamma$, $b\bar{b}$, 
$\tau_h \tau_h$, $4l$, $3l$ and same-sign $2l$. The observed values of signal 
strength modifier $\mu$\footnote{The signal strength modifier $\mu$ multiplies 
the expected SM Higgs boson cross-section in such a way that $\sigma_{observed} 
= \mu \cdot \sigma_{SM}$, so $\mu=\frac{\sigma_{observed}}{\sigma_{SM}}$} and 
the corresponding errors have been reported. While the observed $\mu$ in most 
of these $t\bar{t}H$ channels are more or less consistent with the SM predictions 
within the error bars, the measurement in the same-sign di-lepton channel shows 
an excess of events. The observed signal strength in same-sign $2l$ channel is 
$5.3^{+2.1}_{-1.8}$. Within this category, the same-sign di-muon subsample has the 
largest signal strength, with $\mu=8.5^{+3.3}_{-2.7}$ compared with
$\mu=2.7^{+4.6}_{-4.1}$ for the same-sign di-electron channel and
$\mu=1.8^{+2.5}_{-2.3}$ for the same-sign electron-muon channel.
It is important to note that for this fit, the Higgs boson production
rate other than $t\bar t H$ are considered to be the same to the SM
expectations. We shall discuss later the implications when this condition is
relaxed.
\section{Are these two excesses related to each other?}
These two excesses, reported by CMS collaboration, may seem uncorrelated but one has to keep 
in mind that both the observations are related to the newly discovered bosonic 
state, whose characteristics are yet to be understood completely. If the 
$H\rightarrow \mu\tau$ decay takes place, then it can show up as excesses in different 
final states in $t\bar{t}H$ search. The two excesses seen in data can be the 
two different faces of the same coin and it may contain hints
for BSM physics. In this work, we will try to find whether these two excesses
are related or not.
\section{Scenario at 8 TeV}
To check if the excess in $t\bar{t}H$ same-sign di-muon final state comes 
from the LFV decays of Higgs boson, we have performed a truth-level analysis 
where we have tried to use a similar event selection criteria as used by 
CMS $t\bar{t}H$ analysis, whenever possible with
an equivalent luminosity of 20 fb$^{-1}$, which more or less corresponds to the 
data set recorded by the CMS experiment in 2012. 
We have generated $t\bar{t}H$ events 
for center-of-mass energy 8 TeV using PYTHIA 6 event 
generator~\cite{Sjostrand:2003wg} for two situations, one is when all SM 
decays of Higgs boson are allowed and another is when Higgs boson can decay 
only to a $\mu\tau$ pair. Mass of the Higgs boson and the top quark are taken 
as 125.6 GeV and 173 GeV respectively. The cross-section of $t\bar{t}H$ 
process is considered to be 127 fb~\cite{8TeV}.
Following the CMS analysis, we have selected the events requiring the presence of exactly 
two muons (and no electron) with same sign of charge and at least
four hadronic jets, one of them is required to be a b-quark jet, in the final 
state. Clustering of jets are done using the built-in Pythia module
PYCELL which in turn employs a cone algorithm and incorporates convenient smearing
of the momenta. In PYCELL we granted for an angular coverage of $|\eta|<4.9$ for
the hadron calorimeter with a cell segmentation resembling a generic LHC detector,
i.e., $\Delta\eta\times\Delta\phi=0.1\times 0.1$. In addition, a jet cone of radius
$\Delta R(i,j)=0.5$ has been employed for finding jets.
Both the muons should have transverse momentum ($p_T$) greater than 20 
GeV. Muons and jets should pass the pseudorapidity requirement of $|\eta|<2.4$.
The scalar sum of the $p_T$ of the two leptons and the missing transverse 
energy ($E_T^{miss}$) is required to be above 100 GeV. 
The cuts mentioned above are the event selection cuts from the CMS 
$t\bar{t}H$ analysis, where they have found the following result after applying 
these cuts : the number of events expected from SM $t\bar{t}H$ signal is 
3.1$\pm$0.4 and from SM background is 27.7$\pm$4.7; so the total number of 
events expected from SM in same-sign dimuon channel is 
30.8$\pm$5.1. In this analysis, CMS has observed 41 events, which is an 
excess of 10.2 (5.1) events calculated from the central value (upper edge of error bar) 
of expected number of events. Given that there is an 
excess of events over the SM expectation, the question is if the LFV decay 
of Higgs boson can explain this excess of events or not. One has to keep in 
mind that we have calculated the excess very naively w.r.t the central value 
and upper edge of the 1$\sigma$ error bar and the data is roughly 
consistent with $2\sigma$ error-bar. Thus, the exact number of extra events 
observed in data should not be seen too sacredly. 
Rather the take-away message from this observation is that there is some upward 
fluctuation in data, 
the amount of which is not easy as well as meaningful to quantify without 
doing a sophisticated multivariate analysis, which have been performed by CMS 
as the next step after the cut-based event selection. 
In our analysis we did not apply any multivariate technique, we have used the
event selection cuts from the CMS analysis.
\begin{table}[h]
\vskip 5 pt
\centering
\begin{tabular}{lccc}
\hline
\hline
Process                              &    BR used in    &    $N_{event}$           &    $N_{event}$      \\
                                     &    present      &    present    &    CMS   \\
                                     &    analysis     &    analysis    &   Analysis   \\
\hline
$t\bar{t}H$ $H\rightarrow WW$        &    22.4 \%                 &    2.99                  &    2.4$\pm$0.3 \\
\hline
$t\bar{t}H$ $H\rightarrow \tau\tau$  &    6.3 \%                 &    0.8                   &    0.7$\pm$0.1 \\
\hline
$t\bar{t}H$ $H\rightarrow ZZ$        &    2.8 \%                  &    0.05                  &    0.1$\pm$0.0 \\
\hline
$t\bar{t}W$                                &     -                &    10.2                   &    10.4$\pm$1.5 \\
\hline
\end{tabular}
\caption{Expected number of events after the selection cuts in same-sign dimuon
final state at 8 TeV for the $t\bar{t}H$ production mode for different decay 
channels of Higgs boson.}
\label{table_con_check}
\end{table}
To validate our analysis method, we have checked that the number of events 
obtained by us for $H\rightarrow WW$, $H\rightarrow \tau\tau$, 
$H \rightarrow ZZ$ decay channels in $t\bar{t}H$ production mode and for $t\bar{t}W$ 
background process in the same-sign di-muon final state is fairly consistent with 
the expected number of events reported by CMS after applying the event selection 
cuts. Our observation and numbers from CMS are presented in table~\ref{table_con_check}. 
The branching ratios are taken from~\cite{BR}. We have generated the $t\bar{t}W$ 
process using Madgraph5 event generator~\cite{Alwall:2011uj} at leading order (LO) 
accuracy and later multiplied the event yield by the $K$-factor, which is the ratio 
of cross-sections at NLO and LO. The $K$-factor used for $t\bar{t}W$ process is 1.7. 
The NLO cross-section of $t\bar{t}W$ process is 227 fb at 8 TeV~\cite{Khachatryan:2014ewa}.

We also find the number of events in the same-sign dimuon final state after 
applying the event selection cuts in the $t\bar t H$ production mode at 8 TeV coming 
from $H\rightarrow\mu\tau$ and for branching fraction of 1\% is roughly 0.75.
Now, if $H\rightarrow\mu\tau$ decay is allowed then CMS should also see excess
in same sign $e\mu$ channel. Although, there is no such excess observed but the
uncertainty on the background is quite large (49.3$\pm 5.4$ events). We note in
passing that the same sign electron-muon pair would also show a similar enhancement.
Even 2\% and 3\% branching fractions in this channel can easily be accommodated
given the large uncertainty in the SM background.
From this observation, we can 
conclude that the LFV decay can explain the excess, mostly partially
depending on the quantity of the excess and the branching ratio of LVF decay of 
Higgs boson. We note that although a 2\% branching fraction of the Higgs
boson in the LFV decay channel is not enough to explain the excess in the
same-sign dimuon channel, but it is still a sizeable contribution, nearly
50\% of the SM expectation and brings down the excess within 2$\sigma$ error
of the SM value.
We point out that for 3\% branching ratio of the $H\rightarrow \mu\tau$ 
channel, one obtains roughly 2.3 number of events in the same-sign dimuon final
state. Although higher branching ratios such as these are beyond the constraint 
given by CMS at 95\% confidence level. But we think that it is reasonable to have 
a look at them because of the following two reasons :
\begin{enumerate}
\item In ref~\cite{Harnik:2012pb}, the authors showed that the LFV 
decays of Higgs boson can be sizeable. For example, $H\rightarrow \tau e$ and 
$H\rightarrow \tau \mu$ branching ratios of $\mathcal{O}$(10\%) are 
allowed by low energy constraints coming from $\tau\rightarrow 3\mu$, 
$\tau\rightarrow\mu\gamma$.
\item In the search for LFV decays, it has been assumed that the cross-sections 
of various production processes of Higgs boson are SM-like, which may
not be the case in presence of any new physics. Further, the best fit value of 
the cross-section of gluon fusion production mechanism is $\mu_{ggH}=0.85^{+0.19}_{-0.16}$, 
as reported by CMS \cite{Khachatryan:2014jba}. This is slightly on 
the lower side than the SM expectation, though within uncertainty it is consistent 
with SM prediction. If the production cross-section is lower then the same 
number of observed $H\rightarrow\mu \tau$ events will give rise to a higher value 
of branching ratio. As a result, the upper bound on the branching ratio as reported
by the CMS collaboration would get relaxed. 
For example, we have checked that the constraint
on $\textrm{Br}(H\rightarrow\mu\tau)<1.51\%$ can be further relaxed to 2.19\% (2.85\%)
if one considers the $1\sigma (2\sigma)$ band on the gluon fusion production
as reported by CMS  \cite{Khachatryan:2014jba}. As stated earlier, even this 
branching ratio is not enough to explain the excess but is nonetheless
sizeable and brings down the excess well within 2$\sigma$ error of the SM expected
value. 
Another possibility to boost the branching fraction of $H\rightarrow\mu\tau$ is to
reduce the branching ratio of $H\rightarrow b\bar b$, which reduces the total decay
width of the Higgs boson and subsequently increases the branching fraction in this 
exotic channel. However, one has to respect the experimental bound along with the associated
errors on the LFV process, which are much stringent. 
\end{enumerate}
\begin{figure}[h]
\centering
\includegraphics[width=7cm, angle=-90]{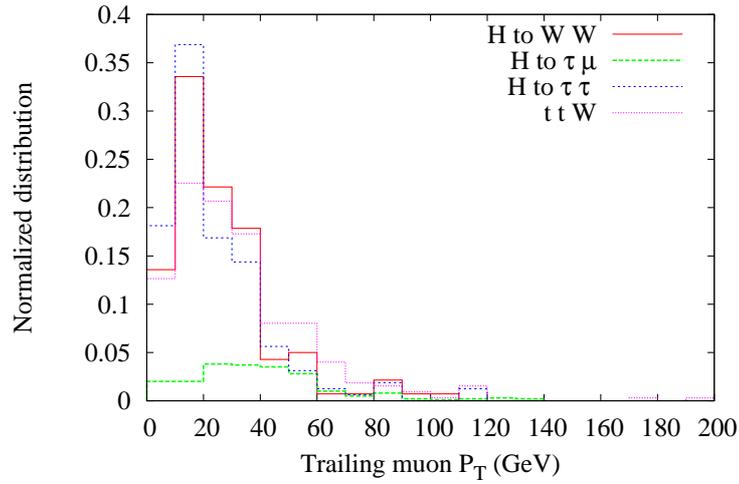}
\caption{Distribution of $p_T$ of the trailing muon in same-sign dimuon final 
state at 8 TeV with respect to the number of events normalised to unity.}
\label{fig:pt_muon2_8TeV}
\end{figure}
\par After the event selection, as one can see from the numbers quoted previously,
the overall yields are still dominated by background events. Therefore, 
CMS analysis uses a multivariate technique after applying the mentioned 
selection cuts. They use Boosted Decision Tree (BDT) which is trained with 
simulated $t\bar{t}H$ signal and $t\bar{t}+jets$ background events, using following six 
input variables : $p_T$ and $|\eta|$ of the of the trailing lepton, the minimal 
angular separation between the trailing lepton and the closest jet, the transverse 
mass of the leading lepton and $E_T^{miss}$, transverse energy of all selected 
jets and leptons ($H_T$) and missing transverse energy of all jets and leptons 
($H_T^{miss}$). Among these input variables, CMS has provided the plot of trailing 
muon $p_T$ in the same-sign di-muon channel and the plot shows that the distribution 
obtained from data is slightly harder than what is expected from the SM. After using 
BDT output as the discriminating variable, CMS has seen a clear excess of events 
in the same-sign di-muon channel which was reported as the observed signal strength
as described previously. 
The CMS paper did not include plots of all six input variables in same-sign 
di-muon final state, so we are unable to compare our findings with observed 
distribution for the other variables except $p_T$ of trailing muon. In fig.~\ref{fig:pt_muon2_8TeV}
we have plotted the $p_T$ distribution of the trailing muon as a function of
the number of events normalised to unity for four different cases - 
(a) SM $H\rightarrow \tau \tau$ (b) SM $H\rightarrow WW$ (c) LFV 
$H\rightarrow \mu \tau$, all of them for the $t\bar{t}H$ production mode and 
the dominant SM background (d) $t\bar{t}W$. We have found that the $p_T$ 
distribution of the trailing muon, obtained from the LFV $H\rightarrow \mu \tau$ 
decay is harder and somewhat similar to the feature that CMS has observed in the 
8 TeV data. This is also expected since the muons emerging from the $H\rightarrow
\mu\tau$ decay are harder compared to the muons coming from $H\rightarrow\tau\tau$.
As a result, $H\rightarrow\mu\tau$ would have a higher efficiency at passing the
CMS selections cuts in the $t\bar t H$ analysis. We also note in passing that
CMS also took into account the contribution from non prompt background
coming mostly from $t\bar t+$jets events, which is also the dominant one.

\section{Future Prospects at 13 TeV}
The run 1 data of CMS is statistically limited for such an analysis involving 
low cross-section signal processes, so the errors on the observed data points 
are huge. However, if both the excesses are genuine then they will be confirmed 
from the run 2 data. We have performed the analysis in 13 TeV in a similar manner 
as described before with an integrated luminosity of 100 fb$^{-1}$.
\begin{table}[h]
\vskip 5 pt
\centering
\begin{tabular}{ccc}
\hline
\hline
Process                              &    Branching Ratio         &    $N_{event}$            \\
\hline
$t\bar{t}H$ $H\rightarrow WW$        &    22.4 \%                 &    24                    \\
\hline
$t\bar{t}H$ $H\rightarrow \tau\tau$  &    6.3 \%                  &    6                      \\
\hline
$t\bar{t}W$                          &     -                      &    60                    \\
\hline
$t\bar{t}H$ $H\rightarrow \mu\tau$   &     2 \%                   &    16                   \\
\hline
\end{tabular}
\caption{Branching ratio of LFV $H\rightarrow\mu\tau$ decay and the corresponding 
number of events obtained in same-sign dimuon final state after applying the event 
selection cuts in the $t\bar{t}H$ production mode at 8 TeV}
\label{table:br}
\end{table}

\begin{figure}[h]
\centering
\includegraphics[width=7cm, angle =-90]{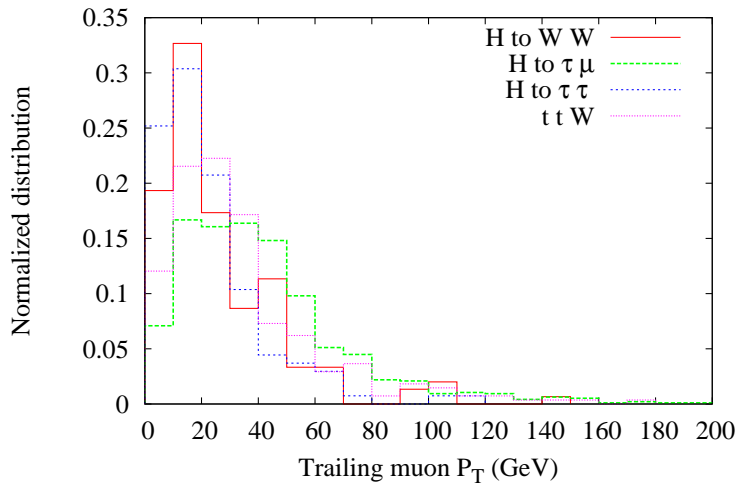}
\caption{Distribution of $p_T$ of trailing muon in same-sign 
dimuon final state at 13 TeV normalised to unity.}
\label{fig_pt_13}
\end{figure}

At 13 TeV the cross-section of $t\bar{t}H$ production mode is taken to be 503 fb~\cite{13TeV}.
We plot the distribution of $p_T$ of trailing muons in same-sign 
di-muon final state as shown in figure \ref{fig_pt_13}. The $p_T$ distribution 
coming from the LFV decay of Higgs boson 
is harder than other SM processes. So we have applied harder $p_T$ cuts on the 
muons, compared to the 8 TeV analysis, in order to distinguish between signal 
and background more efficiently. We require that the leading and trailing muons should pass 
$p_T>50$ GeV and $p_T>30$ cuts respectively. Apart from this change, all other aspects of the 
analysis are same as before. For 13 TeV, the number of events that we have 
obtained from the LFV decay of Higgs boson for the branching ratio of 2\%, and 
number of events obtained from $H\rightarrow \tau\tau$, $H\rightarrow WW$ channels,  
along with dominant SM background $t \bar{t} W$ are reported in table \ref{table:br}.
Just by applying harder $p_T$ cuts, we are not able to kill the background events 
substantially. For a better discrimination between signal and background, one 
should conduct a multivariate analysis which is expected to perform better than 
a cut-based analysis. 
\section{Summary and Outlook}
CMS has observed an excess of events in $t\bar{t}H$ production mode in same-sign 
di-muon final state. Another excess of events in $H\rightarrow \mu\tau$ LFV decay 
channel is also reported by CMS. We have shown that it may be possible that 
these two excesses are correlated and the effect of LFV decay of Higgs boson has 
shown up in the $t\bar{t}H$ search of CMS. 
The branching ratio of LFV decays of Higgs boson are not too much constrained 
from indirect searches performed so far. Keeping that in mind we have explored 
a substantially large range of branching ratio to see how much excess can come 
from this kind of decays of Higgs boson.
However, from the direct search of LFV decay, CMS has given an upper bound on the 
branching ratio BR($H\rightarrow \mu\tau$) which is 1.51\% at 95\% confidence level. 
This bound is obtained 
by assuming the SM-like cross-section for all the production mechanisms of Higgs boson. 
This assumption when relaxed and if the production cross-sections are lower 
than the SM cross-sections (which is still allowed from the most recent measurements 
of CMS collaboration), the upper bound on branching ratio will be pushed 
further in the upward direction. We have checked that the branching
ratio of $H\rightarrow\mu\tau$ can be relaxed upto 3\% if one assumes the production
cross-section through gluon fusion mode is at is lowest value as reported by
CMS. Such a branching is still not enough to explain the excess in the same sign
dimuon channel but can notably bring down the excess of events within 2$\sigma$
value of the SM expectation. 
However, the 8 TeV data is not sufficient to clear all the doubts as it is limited 
by statistics. In the next run of LHC at 13 TeV, we can expect more precise 
measurements of cross-sections and branching ratios of different processes
and will provide a more transparent picture. It is also important to note
that if such LFV decays of the Higgs boson are indeed present then in addition to
the same sign di-muon channel, an excess of similar size would also show up
in the same sign electron-muon final state. Although, the simulation results are
well within experimental uncertainties, however, more data is required to shed
light in this matter.
We also note that the ATLAS collaboration have already searched
for lepton flavour violating Higgs decays to $\mu$-$\tau_{\text{had}}$ final
state in their 8 TeV analysis and did not found any excess~\cite{Aad:2015gha}. 
Nevertheless, it is important to re-examine this issue in the light of 13 TeV run of
the LHC.
\section*{Acknowledgements}
Work of BB is supported by Department of Science and Technology, 
Government of INDIA under the Grant Agreement numbers IFA13-PH-75 
(INSPIRE Faculty Award). SC would like to thank the Council of Scientific and Industrial Research, 
Government of India, for the financial support received as a Senior Research
Fellow and also acknowledges the hospitality of CHEP, IISC during the course
of this work. SC would also like to thank Sourov Roy for helpful discussions.
SM acknowledges the Department of Atomic Energy,
Government of India, for the financial support received as a Senior Research Fellow.
SM also thanks CHEP, IISC because a major part of the work was done there during her 
visit of one month.
%

\end{document}